\documentclass[RNAAS]{aastex631}

\begin{document}

\title{Upper Limits to Long-Term Variability of Solar-Type Stars from Observations of the Open Cluster M67}

\correspondingauthor{Steven Spangler}
\email{steven-spangler@uiowa.edu}

\author[0000-0002-4909-9684]{Steven R. Spangler}
\affiliation{Department of Physics and Astronomy, University of Iowa}

\begin{abstract}
Variations in the luminosity of the Sun on timescales of thousands to millions of years could potentially be responsible for terrestrial climate variations in the Phanerozooic geological period (last 540 million years). Although a few techniques exist for measuring or constraining solar or solar-type stellar variability on long timescales, none have yielded results that are conclusive or reproducible.  In this paper, I consider a method that utilizes observations of an open star cluster with approximately the age of the Sun, specifically M67.  The idea is to measure the width of the main sequence in the region of solar-type stars, here generously defined to be about spectral class G0 - K1. This width gives an estimate of the dispersion in absolute magnitude of nominally solar-type stars. This estimate must be corrected for the presence of undetected binaries and measurement errors in photometric magnitudes. I use the M67 data set of \cite{Geller15}, which provides photometry data and flags known binaries for removal from the sample.  Of the 1278 stars in the data file, 420 were indicated by \cite{Geller15} as single stars (i.e not known binaries) which are members of M67. The sample was further culled by selecting stars with $0.60 \leq (B-V) \leq 0.90$ (approximately spectral class G0 - K1).  It was further restricted to stars with 
$|\Delta m| \leq 0.90$, where $\Delta m$ is the difference between the measured V magnitude of a star, and the predicted value for a star of the same $(B-V)$ value from a theoretical isochrone. This ``culled'' sample consisted of 170 solar-type main sequence stars which are not known to be binaries.  

With this sample, I form an empirical probability density function $p_x(x), x := \Delta m$ and compare it with a theoretical expression from \cite{Spangler25a}.  A satisfactory fit of the theoretical $p_x(x)$ is obtained, with retrieval of two fit parameters.  These are $A$, the fraction of the sample that contains ``residual'' or undetected binaries, and $\sigma$, which is the normalized Gaussian variability of the primary and (if present) secondary star.  In this analysis, I take $\sigma$ to the the quadratic sum of all processes causing the measured brightness of the star to depart from its mean value.  

Statistically-acceptable values of $A$ and $\sigma$ are found to be $0.25 \leq A \leq 0.45$ and $0.10 \leq \sigma \leq 0.135$.  These results do not, however, indicate long-term solar variability at the level of 10 - 14 \%.  The expected value of $\sigma$ from purely photometric noise \citep{Spangler25c}, denoted by $\sigma_{\phi}$ is estimated as $\sigma_{\phi} = 0.101$.  My results are therefore not inconsistent with all of the measured broadening of the main sequence in this data set, other than that due to undetected binaries, being caused by photometric measurement error. In this case, there would be no contribution from intrinsic stellar variability. Generous upper limits to the intrinsic variability contribution to the inferred width are  $\sigma_{IV} \leq 0.058-0.089$.  These limits are not totally devoid of interest in a paleoclimatic context.  However, major improvements in the technique are possible with the use of existing data sets from space astronomy missions such as Gaia and Kepler.  
\end{abstract}

\section{1. Introduction} 
This is the third paper dealing with the width of the main sequence (MS) in an open star cluster, or any sample of main sequence stars.  By the width, I mean the range of apparent or absolute magnitudes measured for the same value of the color, such as (B-V). In a color-magnitude diagram (CMD) of a star cluster, the color allows inference of the stellar photospheric temperature and mass for a main sequence star. The motivation for this investigation is to explore the possibility that low-level variations in the luminosity of the Sun might be responsible for some of climate variability that is seen in the geological record during the past 500 million years.  This period is referred to as the Phanerozoic Eon, and is strictly defined to be the past 539 million years, since the beginning of the Cambrian period \citep{Cohen13}.  

While there is much current attention, popular as well as scientific, on the 1.0 - 1.5K global average warming that has occurred during the past $\simeq$ 150 years, there have been many periods in the Phanerozoic Eon in which the global average temperature was much greater, or much less than the current global average.  The most striking example is the Paleocene-Eocene Thermal Maximum (PETM), in which mean global temperatures were 5 - 8 K higher than the present global mean \citep{McInerney11}.  This event, and others in the geological record, were obviously not caused by anthropogenic CO$_2$ enhancement \footnote{It is necessary to point out that the prevailing opinion among geologists is that the PETM was caused by enhanced atmospheric CO$_2$, but generated by natural processes such as volcanism \citep{McInerney11}.}. As will be discussed in Section 6 of this paper, even variations in the solar luminosity of one to a few percent could cause temperature variations of 1 to a few degrees K, without the agency of geophysical feedback processes. 

The present paper addresses the possibility of constraining long-term solar variability through study of solar-type stars.  It is axiomatic that many stars in the Galaxy are very similar to the Sun, i.e. spectral class G main sequence stars.  By studying such stars, either individually over a long time span, or in statistically valid samples at a single time, it might be possible to determine if they, and by extension the Sun, manifest variability at the level of one to several percent over long timescales.  

The term ``Long Term Variability'' is deliberately left vague.  What I have in mind is variations on timescales of centuries to millenia, which might account for climate variations in the historical and archaeological record, or on timescales of millions of years, which might be responsible for events in the geological record.  In the next section, I will discuss how various techniques probe the different potential timescales.  

To limit the scope of this paper to a manageable one, I will not discuss the extensive efforts to make observations of the Sun which would indicate long-term variability.  Those efforts include direct measurements of the Total Solar Irradiance (TSI, essentially the solar Poynting Flux at the orbit of Earth) since spacecraft measurements became available in the 1970s, and efforts to infer TSI over much longer timescales baselines by proxies, most commonly the abundance of radiogenic isotopes of carbon and berylium.  Good entry points to the literature here would be \cite{DudokdeWit17} and \cite{Usoskin23}, respectively.  

\subsection{1.1 Previous Efforts to Constrain Long-Term Solar Variability from Observations of Solar-Type Stars}

The approach employed in this paper is to study a sample of solar-type stars to see if they show evidence of low level variability. This involves solar ``Doppelgangers'' consisting of main sequence stars of spectral classes from late F to early K.  The assumption then would be that the Sun itself probably displays similar behavior.  One way to do this is to directly measure the flux from the solar-type stars over as long a time baseline as possible.  This has been done previously by \cite{Radick18} and \cite{Judge20}.  In \cite{Radick18}, measurements were made of a quantity known by the arcane name of $R^{'}_{HK}$, and which is the normalized flux in the cores of the H and K lines of calcium.  In the case of the Sun, this emission is due to processes in the chromosphere, and varies during the solar cycle. \cite{Radick18}  also measured variations in the continuum flux in two bandpasses in the visual part of the spectrum (the Str\"{o}mgren b and y filters). The data set analysed had a record length of 24 years. The emphasis in \cite{Radick18} was on the relationship between emission line variations (chromospheric processes) and continuum variability (photospheric processes), with the goal of better understanding magnetic cycles in solar-type stars. However, magnetic cycles in the Sun and solar-type stars could well be related to long-term variability \citep{Lydon95}.

\cite{Judge20} used the same continuum flux data set as \cite{Radick18} for the same set of stars, but directly addressed the question of long-term variability.  The title of \cite{Judge20} clearly shows the relevance to the present investigation. The mean duration of the time series for each of the 22 stars in a subset of the \cite{Radick18} sample was 17 years.  \cite{Judge20} found, in their sample of 22 solar analogs, a number of stars which had linear trends in brightness that would be climatologically significant if continued over a period of $\sim 250$ years (the time to the Maunder Minimum). Quantitatively, \cite{Judge20} place limits on linear trends in their sample as those which would produce radiative forcing changes of $\pm 4.5$ W/m$^2$ over 250 years, as compared to $+2.2 \pm 1.1$ W/m$^2$ as the estimated global forcing due to anthropogenic CO$_2$ forcing since the end of the 19th century \citep{Judge20}. If such trends for solar-type stars continued over longer periods of millenia or tens of millenia, the changes would obviously be even more significant.  

While work like that reported by \cite{Radick18} and \cite{Judge20} arguably represents the most effective way of addressing the concern of this paper, such studies are limited by the relatively short time span over which such measurements, carried out with the same instrument and same calibration approaches, can be realized.  In \cite{Radick18} and \cite{Judge20} the total record lengths were of the order 2-3 decades.  

\subsection{1.2 Measurements of Solar Type Stars in Open Star Clusters} 
A complementary approach to the investigations of \cite{Radick18} and \cite{Judge20} is to study solar-type stars in open star clusters.  Open star clusters are composed of stars of the same distance, age, and chemical composition.  As a result, the Russell-Vogt theorem \citep{Abell66} indicates that all main sequence stars of the same mass should have the same luminosity.  If they do not, it could be an indicator of long term variations. 

The basic assumption, previously clearly stated by \cite{Giampapa06} and \cite{Curtis17}, is that solar-type stars in an open star cluster could enter and leave periods of variability on timescales much shorter than their ages.  Each star in this ensemble would then be varying independently of the others, and the distribution of luminosity or absolute magnitude would be a measure of their variability amplitudes.   

A difficulty with using open star clusters in this capacity is that very few clusters persist for periods of time equal to the age of the Sun. Indeed, \cite{Hunt23,Hunt24} report that only a small percentage of the thousands of open clusters they investigated were older than $10^9$ years. Favorites of amateur astronomers such as M44, the Hyades, and M35 are a few hundred million years old, so the solar-type stars in those clusters resemble the Sun in the time of the early solar system.  It is well known that solar-type stars with ages well under $10^9$ years rotate faster, and have more variability and activity than similar stars of age several billion years. An analysis of this sort utilizing open star clusters must therefore concentrate on the few that have ages comparable to the Sun.  Probably the best choice in this respect is M67. In this paper, I will adopt an age of 4.0 Gyr for M67, which is close enough to that of the Sun for the stars in M67 to be considered ``solar-like''.  Estimates for the age of M67 in the literature range from about 3.3 Gyr to the age of the Sun.  

The idea of using M67 to investigate the possibility of long term variability of solar-type stars is not original with this paper.  \cite{Giampapa06} measured the $R^{'}_{HK}$ value in solar-type stars in M67, and noted a number of stars in which a low value was measured.  \cite{Giampapa06} used $R^{'}_{HK}$ as a proxy for stellar luminosity, and concluded that a number of the M67, solar-type stars were in a Maunder Minimum state. This result was disputed by \cite{Curtis17}, who measured interstellar absorption in the H and K lines of calcium for non-solar stars in M67, and concluded that the low values of $R^{'}_{HK}$ reported by \cite{Giampapa06} were due to interstellar extinction in the direction of M67. 

\subsection{1.3 Scope of the Present Investigation} 
In this paper, I also use the solar-type stars in M67 as a sample with which to measure or constrain the amplitude of long-term luminosity variation.  The approach I use is to measure the width of the main sequence in the solar part of the Hertzsprung-Russell (HR) diagram.  The fundamental observable I use is defined in the notation of \cite{Spangler25a} by $x$, 
\begin{equation}
x := \Delta m = m_0 - m
\end{equation}
where $m$ is the measured apparent magnitude of a star, and $m_0$ is the expected apparent magnitude of a star with the same measured color, as determined by a theoretical isochrone. Equation (1) is defined such that $x > 0$ if the star is brighter than the theoretical prediction.  Obviously, the absolute rather than apparent magnitudes could be used as well. In the perfect case of noise-free measurements of a cluster with all stars being single, the dispersion of this quantity would be directly proportional to luminosity variations. 

In reality, there are at least four other mechanisms for broadening the main sequence, even in the case of nonexistent luminosity variations. 
\begin{enumerate}
\item A large percentage of the stars in an open cluster will be binaries, or multiple star systems, with the blended light being brighter than that of the primary alone.  Previous studies of open star clusters have shown that the fraction of stars that are binaries (or multiples) ranges from less than 20 \% to 50 \% or more \citep{Geller15,Childs24,Childs25}, and the percentage of binaries is found to be higher in the cluster core than the periphery for some clusters \citep{Childs24,Motherway24}. The approach I favor is to choose a cluster for which many of the binaries have been detected, and thus can be removed or ``culled'' from the sample for analysis.  The distribution function of secondary masses can be modeled, and the resultant probability density function (pdf) of $x$, $p_x(x)$ modeled and fit to an observed distribution. 

The motivation for applying or fitting a theoretical $p_x(x)$ to a culled sample rather than the initial sample of member stars is that a model function will inevitably have inaccuracies due to mathematical and physical approximation.  It should be the case that the smaller the number of ``residual'' binaries, the smaller the impact on the retrieved parameters of the fit $p_x(x)$ function.   

In \cite{Spangler25a} (more complete presentation in \cite{Spangler25b}), an expression was derived for $p_x(x)$ which included an assumed distribution function for the masses of the secondary stars, and a Gaussian variability function which described variations in the luminosity of the primary and secondary stars. 

\item If the HR diagram of a star cluster is plotted as a color-magnitude diagram, errors in the magnitude measurements in both colors will affect the width of the HR diagram.  In \cite{Spangler25c} (more complete presentation in \cite{Spangler25d}), it was found that the width so introduced was surprising large, although simple arguments for understanding this were presented.  The effect of noise errors on the dispersion in $x$ was characterized by a Gaussian standard deviation $\tilde{\sigma}$ given by Equation (7) of \cite{Spangler25c}. An assumption I will employ in this paper is that the $x$ distribution for a real star cluster may be considered the convolution of a noise Gaussian characterized by $\tilde{\sigma}$ with the pdf of a distribution caused by binary contamination and stellar variability presented in \cite{Spangler25a}. 

\item Differential extinction across the face of a cluster will inevitably occur at some level, due to patchiness in the interstellar medium. M67, being relatively nearby (890 pc) and at a Galactic latitude of about 32$^{\circ}$ should be much less affected by this than many other clusters, but the effect will still be present at some level. As noted above, \cite{Curtis17} argued that interstellar absorption was responsible for the low end of the H \& K line intensities reported by \cite{Giampapa06}, but did not provide estimates of star-to-star variations (differential extinction). \cite{Childs25} provide an extinction map for the vicinity of M67, which does reveal some spatial structure.  

\item Since open star clusters are of finite size, not all member stars are at the same distance, and this will result in dispersion in $x$ due to the inverse square law.  To first order, this can be calculated in a straightforward manner given the measured core radius of a cluster, which is available for many clusters \citep{Hunt23,Hunt24,Childs24,Childs25}.  A complication could arise because some clusters, like the Hyades and M44, exhibit ``tidal tails'' aligned with the L1 and L2 points as part of the disruption process \citep{Cantat22}. This will cause the star cluster to depart from spherical symmetry, and be extended along an axis.        
\end{enumerate}   
In this paper, I will correct for broadening mechanisms \# 1 and \# 2 above. Mechanism \#3, which may be important, would require extinction measurements for the line of sight to each star considered.   Such data are not available at present, although I hope to address this point in a future investigation.  Mechanism \#4 will produce fluctuations smaller than those observed, but should also be considered at greater length in a future study.  

\section{2. Data Selection}

The data utilized were from an online file made available in \cite{Geller15}.  Values of V and (B-V) magnitudes in that file were taken from the measurements by \cite{Montgomery93}.  The appeal of this data set, other than its ready availability in machine-readable form, was the secure establishment of M67 membership on the basis of radial velocity, and the specification of stars as being member binaries or member single stars. The radial velocities used for both membership and binary properties were acquired as part of the University of Wisconsin program of open cluster spectroscopy, which has been in progress for many years.  

\subsection{2.1 Stellar Sample Size}
My approach was to select stars on the basis of being a single star member of M67, then plotting them on a standard color-magnitude diagram as a plot of $m_V$ (apparent magnitude in Johnson V band) versus $(B-V):= m_B - m_V$.  The total number of stars selected was 414.  

The obvious fact should be recognized that what I refer to as ``single stars'' are in fact ``stars not known to be binaries''.  In other words, \cite{Geller15} have identified spectroscopic binaries in M67, which we discard in the selection process, but there are other binaries which were not detected by \cite{Geller15}, primarily because their periods are longer than the duration of the University of Wisconsin program. This is directly discussed by \cite{Geller15}, who provide an estimate of the total binary fraction, as well as the fraction that are detected.  These latent binaries will be referred to by the obvious name ``undetected binaries''.  An undetected binary will be brighter than a true single star.  

The HR diagram formed from these 414 nominally single stars is shown in Figure 1. The presence of a ``binary sequence'' 0.75 magnitudes above the main sequence is an obvious indicator of some of the undetected binaries.   
\begin{figure}[h]
\begin{center}
\includegraphics[scale=0.50,angle=0]{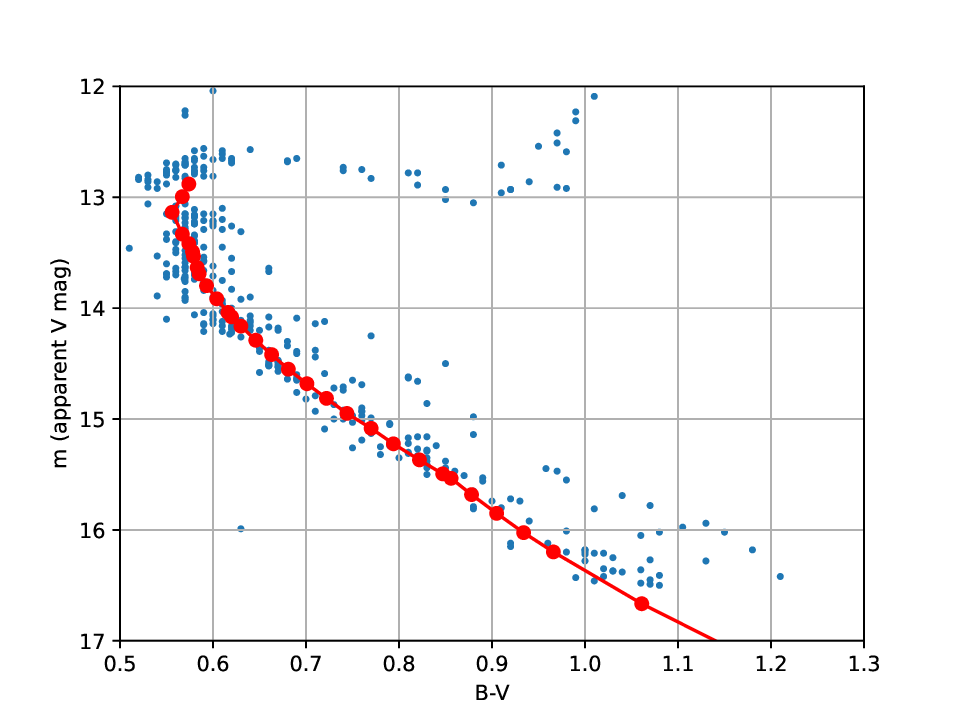}
\caption{The Hertzsprung-Russell diagram for all ``single'' stars in M67 (NGC2682).  The abscissa is $(B-V)$ and the ordinate is $m_V$. Blue data points represent data from \cite{Geller15} which are known to be cluster members, and not demonstrated to be binaries. The solid red curve is a theoretical isochrone for a cluster with the age and chemical composition of M67, and nominally corrected for distance, absorption, and reddening.}
\end{center}
\end{figure}

\subsection{2.2 Theoretical Isochrones}
My analysis requires a curve in the color magnitude diagram that represents the locus of points for the single stars, and defines $m_0(B-V)$, to use the variable of Equation (1).  Although a purely empirical fit to the data points in the HR diagram would be acceptable, it seems preferable to utilize a physically-based curve based on a stellar evolution calculation.  

For this purpose, I used the code discussed in \cite{Marigo08}, which may be run and the results collected via an online site\footnote{The URL for the control panel of the code is at: \url{http://stev.oapd.inaf.it/cgi-bin/cmd}.  This service is maintained by Leo Girardi at the Osservatorio Astronomico di Padova. }.  I used this online tool because it was also used by \cite{Geller15}. 
The code requires inputs of age and metallicity, for which I chose $T=4.0$    Gyr and solar metallicity. Using it as a reference curve on a color-magnitude diagram requires also specifying the distance, reddening $E(B-V)$, and $R$, ratio of total to selective absorption. For these I chose $d = 889$ pc, $E(B-V) = 0.030$, and $R = 3.1$.  The choice of $E(B-V)$ was taken from \cite{Geller15}, who gave a range of estimates of $E(B-V)$  as $0.015 \leq E(B-V) \leq 0.056$. \cite{Geller15} used this curve only for purposes of ``guiding the eye'', whereas I use it for the reference function $m_0(B-V)$.  This isochrone is shown in Figure 1 as a solid red curve. 

\subsection{2.3 Further Filtering of Stellar Sample} 
The goal of this project is to study solar-type stars.  Figure 1 shows additional stars at the main sequence turnoff (more massive than the Sun), subgiants, and red dwarfs.  To restrict the sample to solar-type stars and solar analogs, I further selected stars in the sample which had $0.60 \leq (B-V) \leq 0.90$, and $13.0 \leq m_V \leq 16.0$.  This selection includes main sequence stars between about spectral classes G0 and K1. 

\section{3. Calculation of the Empirical $p_x(x)$ Function} 
For each star in this windowed sample of nominally solar-type stars, the quantity $x$ as defined in Equation (1) was calculated, with $m_0$ being the apparent magnitude of a star on the theoretical isochrone with the same value of $(B-V)$ as that measured, and $m$ being the measured $V$ band magnitude from \cite{Geller15}. The value of $m_0$ for each star was determined by linear interpolation between the points returned by the model, indicated in Figure 1 by the red points.  A final filtering operation consisted of discarding the few stars with $x$ outside the range $-0.40 \leq x \leq 0.90$, on the assumption that such stars could not have their brightness anomalies determined by binarity, measurement error, and natural variability.  The final sample of ``solar-type'' main sequence stars for analysis consisted of 170 stars.   

The sample of 170 stars was then collected into a histogram in $x$, with bin size $\Delta x$ chosen by the program.  For the results presented here, a value of $\Delta x = 0.050$ was chosen as a compromise between resolution and adequate statistics.  The raw histogram was converted into a normalized probability density $p_x(x)$ via the formula
\begin{equation}
p_x(x_i) = \frac{\Delta N_i}{N_{\ast} \Delta x}
\end{equation}
where $x_i$ is the value of $x$ in the ith histogram bin, $\Delta N_i$ is the number of stars in the ith bin, and $N_{\ast}$ is the total number of stars in the sample (170). The empirical $p_x(x)$ so calculated is represented in Figure 2 by the blue data points. 

The errors on the $p_x(x)$ values are calculated assuming a Poisson distribution, so
\begin{equation}
\frac{\sigma_p}{p_x(x_i)} = \frac{1}{\sqrt{\Delta N_i}}
\end{equation}
For bins outside the main part of the distribution, $\Delta N_i$ was in some cases $\leq 2$.  In these cases, $N=2$ was used in Equation (3).  
\begin{figure}[h]
\begin{center}
\includegraphics[scale=0.50,angle=0]{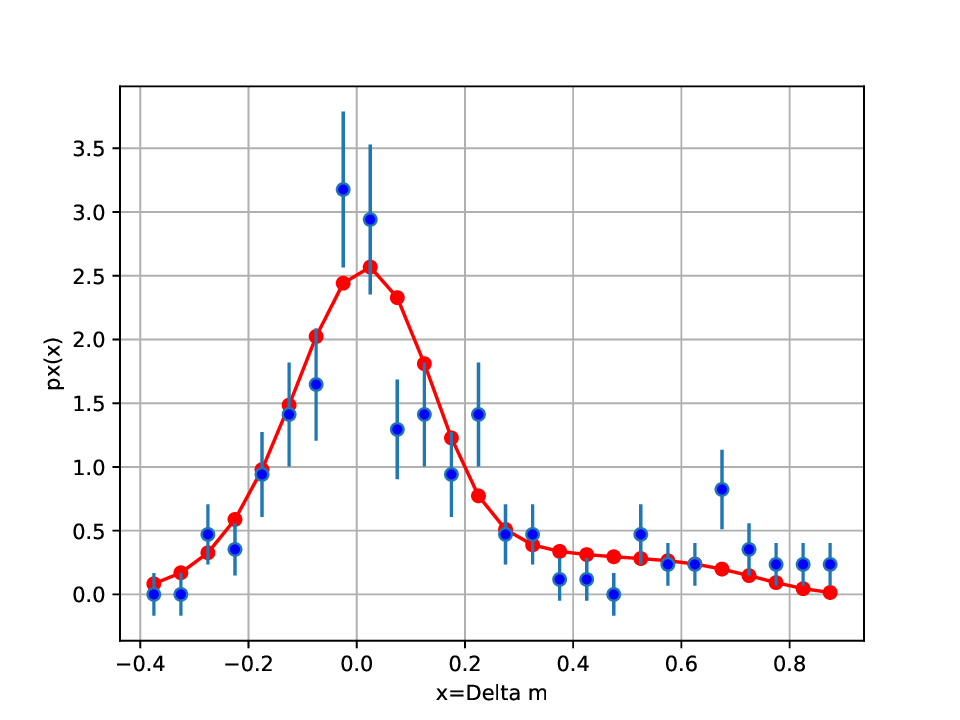}
\caption{The empirical probability density function $p_x(x)$ for the solar-type part of the main sequence for M67 (blue data points). The red curve with solid red guide points gives a fit of a theoretical probability density function $p_x(x)$ to the measured function for M67.  The model incorporates a Gaussian width $\sigma=0.117$ and a binary fraction $A=0.353$ (defined in Section 4).}
\end{center}
\end{figure}

The data in Figure 2 show the form of $p_x(x)$ expected from the discussion of \cite{Spangler25a}; a roughly Gaussian core centered close to 0, and an asymmetric tail extending for positive values of $x$ out to $x \simeq 0.8$.  The tail is interpreted as undetected binaries.  The red curve in Figure 2 represents a fit of the theoretical model obtained in \cite{Spangler25a}, and discussed further in Section 4 below.  

\subsection{3.1 Search for Trends in Empirical $p_x(x)$ Relationship}
The purpose of the theoretical isochrone is to approximate the locus of points for single main sequence stars as a function of mass or color.  For the purposes of this paper, the theoretical isochrone is a physically-motivated fitting curve to the lower envelope of the color-magnitude diagram (CMD).  An offset between the theoretical isochrone and the true envelope is expected and acceptable, and can be removed as a simple fitting parameter.  However, an offset (between the true locus of single star points and the isochrone) which varies with $(B-V)$ could be problematic, since it would artificially broaden the empirical distribution, especially the Gaussian core of the distribution which contains information on possible stellar variability.  

To investigate this point, and to be at least aware of the magnitude of the problem should it be detected, the following procedure was undertaken.  
\begin{enumerate}
\item The individual $x_i$ values for all $i \in {1,N_{\ast}}$ were plotted as a function of $(B-V)_i$.
\item Since the fit to the whole data set, with $x \leq 0.90$ would have been affected by the undetected and unremoved binaries, a ``clipped'' subsample with $|x| \leq 0.30$ was formed. A plot of $x_i$ as a function of $(B-V)_i$ for the clipped sample is shown in Figure 3.
\item A linear regression was made to the clipped data set (Python subroutine scipy.stats.linregress). 
\end{enumerate}
\begin{figure}[h]
\begin{center}
\includegraphics[scale=0.50,angle=0]{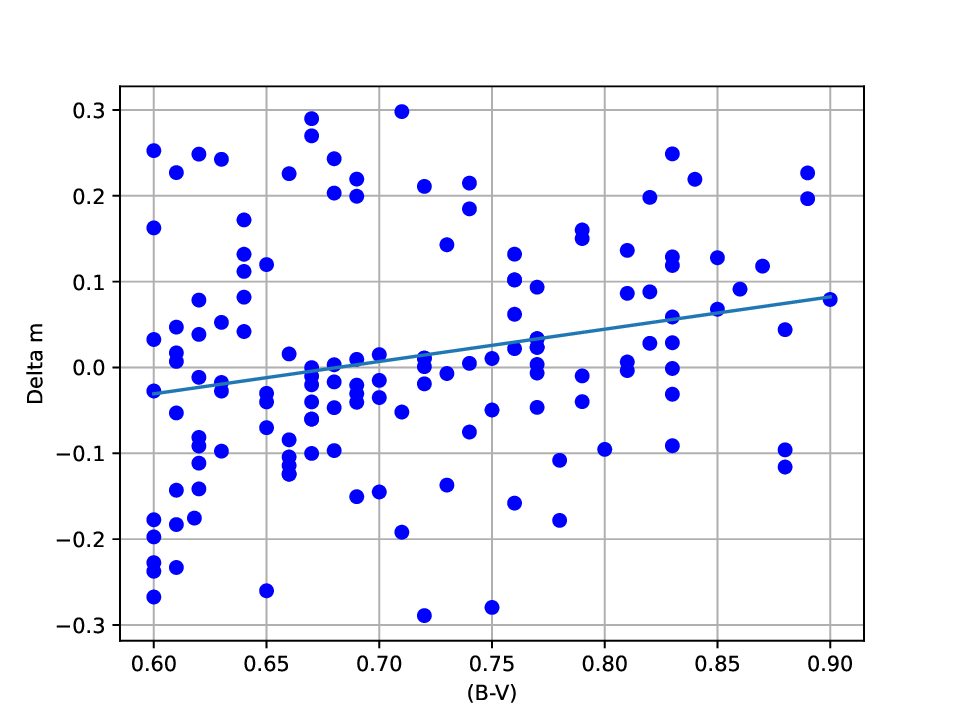}
\caption{Residual magnitudes $x=\Delta m$ values for stars with $|x_i| \leq 0.30$ in the sample, plotted versus color $(B-V)$.  The purpose of this plot is to look for possible trends.  The line is a linear regression to the sample, and possesses a slope of 0.376 magnitudes of $V$ per magnitude of $(B-V)$
.}
\end{center}
\end{figure}
Visual examination of the plot does not reveal major evidence for a shift in the $x$ distribution with $(B-V)$ for $0.60 \leq (B-V) \leq 0.90$.  A linear fit of the form $x = a(B-V) = b$ yields a slope $a = 0.376 \pm 0.132$ and intercept $b = -0.256 \pm 0.095$ (shown by the solid line). The data in Figure 3 are therefore marginally consistent (i.e. the slope differs from zero at slightly less than the 3 $\sigma$ level) with a linear trend over the color range of interest. The nominal linear trend shown in Figure 3 would produce a change of $x =  0.113$ over the range in $(B-V)$ that is considered.  

It was decided the best approach to estimating the effect of this trend (if present) was to carry out the fitting analysis of the theoretical model for $p_x(x)$ to the data both without and with this trend removed.  This analysis will be presented in Sections 4.1 and 4.2, respectively.

\section{4. Fitting of Theoretical $p_x(x)$ Function to the Empirical Function}

This section will be concerned with fitting the theoretical $p_x(x)$ function derived in \cite{Spangler25a,Spangler25b} to the empirical function shown in Figure 2. The function in \cite{Spangler25a,Spangler25b} was derived in 2 forms, a semi-analytic expression and a fully-analytic one.  The semi-analytic form was completely analytic except for a convolution step which was done numerically.  The second form was fully analytic, but adopted an approximation for the contribution of light from the secondary in binary members of the sample.  The expressions presented in \cite{Spangler25a,Spangler25b} assume that intrinsic variations of both primary and secondary components of binaries (and of single stars) are Gaussian-distributed, and that the pdf of the mass ratio $q$ is uniformly distributed between 0 and 1.00 (Equations (13) and (14) of \cite{Spangler25b}).

I used the semi-analytic expression in the fit to the data of Figure 2.  The theoretical expression is determined by two parameters, $\sigma$ and $A$.  The parameter $\sigma$ is the Gaussian rms of intrinsic luminosity variations in the primary and secondary.  $A$ is the fraction of undetected binaries in the sample of stars.  In addition, there is a constant offset parameter (in $x$) $x_{off}$, presumed small, between the theoretical isochrone used and the peak of the Gaussian central distribution. This can also be determined in the fitting process, and was confirmed to be small. The goal of the analysis is to find which values of $\sigma$ and $A$, if any, satisfactorily describe the observed pdf, and by extension, the observed cluster HR diagram.  

Given the goal of this study, $\sigma$ is the quantity of primary interest, since it includes the magnitude of long-term luminosity variability of these solar-type stars.  $A$ is a necessary fitting parameter, which must be solved for together with $\sigma$, but is also of interest in its own right.  

\subsection{4.1 Fit to the Entire $p_x(x)$ Function}
The main analysis consists of fitting the entire $p_x(x)$ with $-0.40 \leq x \leq 0.90$ to adequately model undetected binaries and obtain a global fit.  I use the data set without the linear trend (Section 3.1) removed, since linear regression for the whole sample would be biased by the undetected binaries.  

The parameter chosen to measure the goodness of fit is the reduced chi-square parameter $\chi^2_{\nu}$ \citep{Bevington69}. In this case, $\nu$, the number of degrees of freedom in the fit is equal to the number of data points in Figure 2  minus the number of fit parameters, $(\sigma, A, x_{off})$ or 23. For a good fit, $\chi^2_{\nu} \simeq 1$.  For a small number of degrees of freedom, values of $\chi^2_{\nu}$ significantly in excess of unity can be acceptable, but as the number of degrees of freedom increases, values much larger than unity indicate that the model function is an improbable representation of the underlying process.  
 
A grid seach was carried out over $A$ and $\sigma$, and to a lesser extent, $x_{off}$.  The fit parameters $A$ and $\sigma$ did not seem to be strongly correlated, so I searched for the acceptable ranges of these parameters by orthogonal slices in the $(\sigma, A)$ plane.  

Figure 4 shows a plot of $\chi^2_{\nu}$ as a function of $A$ for a fixed value of $\sigma = 0.117$ and $x_{off}=-0.010$.  The horizontal lines represent $\chi^2_{\nu}$ values corresponding to probabilities of 5\%, 2 \%, and 1 \% in order from lowest to highest. These values are taken from \citep[][Table C4]{Bevington69}. The solid plotted points represent values of $\chi^2_{\nu}$ calculated at discrete values of $A$.  The continuous curve is a quadratic fit to those points. 

\begin{figure}[h]
\begin{center}
\includegraphics[scale=0.50,angle=0]{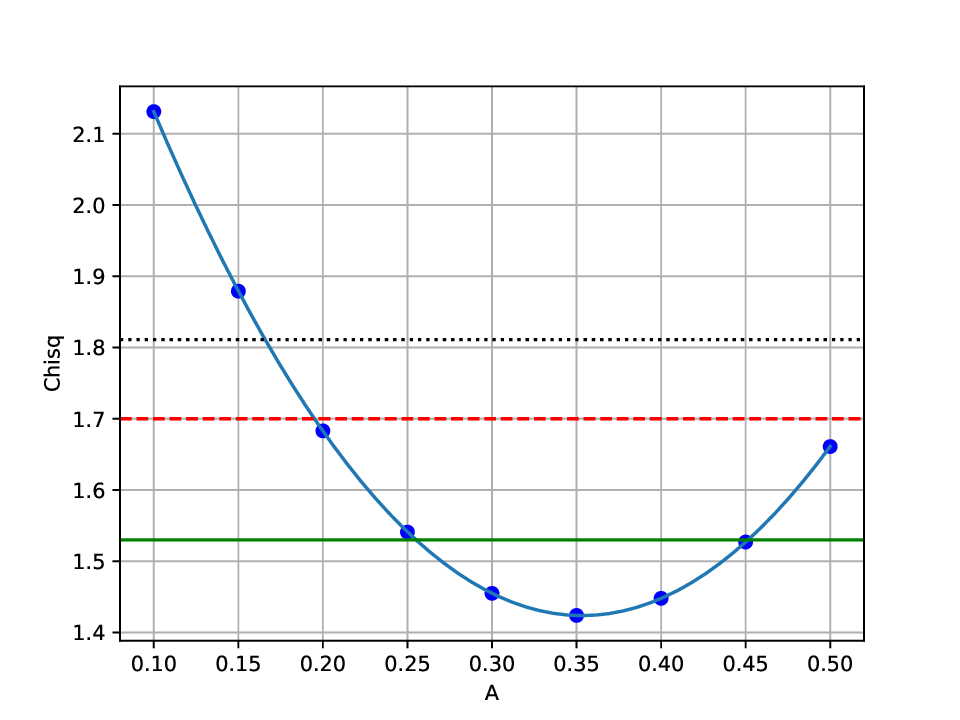}
\caption{Slice through $\chi^2_{\nu}$ space for fit to empirical $p_x(x)$ function.  This is a slice for fixed $\sigma=0.117$, as a function of the binary fraction $A$. The arbitrary offset parameter $x_{off}=-0.010$. The green, red-dashed, and black-dotted lines correspond to $\chi^2_{\nu} = 1.530, 1.700, 1.811$, respectively, which represent probabilities of 5\%, 2\%, and 1\%.}
\end{center}
\end{figure}

The theoretical $p_x(x)$ function did not result in very good fits to the data, but I will adopt a 5 \% probability threshold as acceptable in the following discussion. The range of $A$ which is acceptable, given the above criterion, is $0.25 \leq A \leq 0.45$. This result is given in Table 1. Further comment on the feasibility of this range is given in Section 5.1.    

The corresponding slice in $\chi^2_{\nu}$ space of $\chi^2_{\nu}$ as a function of $\sigma$ for fixed $A = 0.30$ (again, $x_{off}=-0.010$) is shown in Figure 5. The horizontal lines for fixed values of $\chi^2_{\nu}$ have the same significance as in Figure 4. The continuous curve is a cubic fit to the $\chi^2_{\nu}$ values at discrete choices of $\sigma$.  The reason for choosing the slice at $A=0.30$ instead of the minimum in Figure 4, $A=0.35$, is discussed in Section 5.  

\begin{figure}[h]
\begin{center}
\includegraphics[scale=0.50,angle=0]{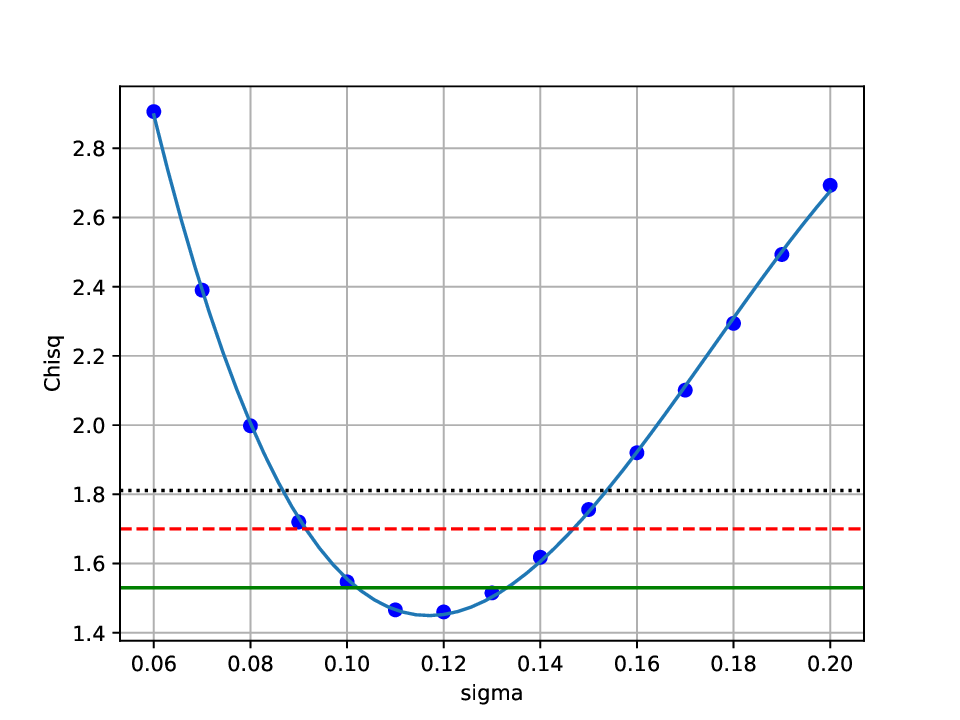}
\caption{Slice through $\chi^2_{\nu}$ space for fit to empirical $p_x(x)$ function.  This is a slice for fixed $A=0.30$, as a function of the Gaussian width $\sigma$.  The arbitrary offset parameter $x_{off}=-0.010$. The green, red-dashed, and black-dotted lines have the same significance as in Figure 4.}
\end{center}
\end{figure}

Applying the same criterion of statistical acceptability as used for the parameter $A$, I find an acceptable range of $\sigma$ to be $0.10 \leq \sigma \leq 0.135$. This result is also shown in Table 1.  

The optimum model curve for $p_x(x)$ with parameters $A=0.353, \sigma=0.117, x_{off}=-0.010$ is given by the red solid curve in Figure 2.  I judge this to be an acceptable representation for the extended ``tail'' to the distribution, particularly for $x \geq 0.20$, as well as the core of the distribution.  

\subsection{4.2 Restricted Fit to Gaussian ``Core'' of the Distribution}
I undertook a restricted fit of the theoretical $p_x(x)$ to the central ``core'' of the distribution function, with $-0.30 \leq x \leq 0.30$.  This was done for two reasons.
\begin{enumerate}
\item To determine the effect of the trend removal discussed in Section 3.1 and illustrated in Figure 3. 
\item To possibly obtain an improved value of $\sigma$, the parameter of primary interest to this study, by concentrating on the core of the distribution, which contains such information.  For these purposes, the binarity parameter $A$ was fixed by the analysis in the preceding section, on the grounds that a fit to the entire distribution is better in determining it.  
\end{enumerate}

I therefore made a second fit of the theoretical $p_x(x)$ function to the data set, but the variable was now a ``detrended'' version of $x$.
\begin{equation}
x' := x - a(B-V) -b
\end{equation} 
These are values of $x$  corrected for the linear regression to the entire data sample shown in Figure 3. The variables $a$ and $b$ in Equation (4) are the slope and intercept of the fit line. This fit adopted a fixed value of $A = 0.350$ taken from the fit to the entire data sample, as described in Section 4.1 above.  

The fits to the detrended sample yielded a range in the parameter $\sigma$ of $0.085 \leq \sigma \leq 0.115$, with a best-fit value of $\sigma = 0.100$. These results are also given in Table 1. This range is comparable to that cited in Section 4.1 above, although slightly lower, as expected given the detrending operation.  For the remainder of the analysis, I will consider the acceptable range of $\sigma$ to be that given in Section 4.1, $0.10 \leq \sigma \leq 0.135$, with the recognition that the slightly smaller range from the detrended sample is also worth consideration. \\

{\bf Table 1. Fit of $p_x(x)$ to Data Set } \\ 
\begin{tabular}{||l|l|l||} \hline \hline
Fit Parameter & Best-Fit Value & Acceptable Range \\ \hline
$A$ (binary coefficient)  & 0.353 & $0.25 \leq A \leq 0.45$ \\ \hline
$\sigma$ (width of core)  & 0.117 & $0.100 \leq \sigma \leq 0.135$ \\ \hline
$\sigma$ (detrended, central core) & 0.100 &   $0.085 \leq \sigma \leq 0.115$ \\ \hline
\end{tabular} \\
\section{5. Discussion of Parameters Retrieved from Fits}
The model function described in \cite{Spangler25a} possesses two parameters, as discussed in Section 4, $A$ and $\sigma$.  There is also an ``offset parameter'' $x_{off}$ which is expected to be close to zero, and that expectation was verified in the fitting process.  In this section, I discuss the significance of the parameters that have emerged from the fitting analysis. 
\subsection{5.1 Binary Fraction $A$}
The result on $A$ discussed in Section 4.1 and given in Table 1 is surprisingly, and perhaps implausibly high, given that the sample analysed is ``culled'', meaning that known binaries have been removed.  For this reason, I believe only the lower part of the acceptable fit range given in Table 1 is viable, and the upper part is incompatible with independent information.  The reasons for this conclusion are given below. 

The sample of M67 stars given in \cite{Geller15} had 562 members, chosen on the basis of radial velocity and proper motions.  Of these, 142 were demonstrated to be binary or multiple star systems on the basis of radial velocity measurements carried out over several decades. This left a sample of 420 ``single'' stars, of which 414 passed the selection algorithm in my program, and which provided the sample studied in this paper. However, as emphasized in \cite{Geller15} and repeatedly noted here, some fraction of these 414 stars will still be binaries that eluded detection in the investigation of \cite{Geller15}. The variable $A$ which is fit for in my investigation is the fraction of remaining or residual binaries in the main sequence subsample of stars. 

The binary fraction detected by \cite{Geller15}  is $\frac{142}{562} = 0.253$. \cite{Geller15} estimate that the total binary fraction, including those undetected by radial velocity variations, is $0.34 \pm 0.03$.  In what follows, I form an alternative estimate of the total (unculled) binary fraction, given a value for $A$, which will be assumed to be correct.  

Let $N_s$ be the total number of stars in a sample, including binaries and single stars.  In the present case, this would be the number of M67 member stars.  Let $f_T$ be the true fraction of stars that are binaries or multiples, and $f_o$ be the fraction that are determined to be binaries by independent means, such as radial velocity measurements.  

I define $N_{sa}$ to be the number of stars in the culled sample, $N_{sa} = N_s(1-f_o)$.  If $N_{bu}$ is the number of undetected or residual binaries in the culled sample, then my fit parameter $A$ is
\begin{equation}
A = \frac{N_{bu}}{N_{sa}} = \frac{f_TN_s - f_oN_s}{N_s(1-f_o)} = \frac{f_T - f_o}{1- f_o}
\end{equation}
Given a known value of $f_o$ and a fit value (or range of acceptable values) of $A$, Equation (5) allows one to estimate $f_T$.  
\begin{equation}
f_T = A + f_o(1-A)
\end{equation}
Using the values of minimum acceptable, best-fit, and maximum acceptable values of $A$ from Table 1 ($A = 0.250, 0.350, 0.45$, respectively) I obtain values of $f_T = 0.44, 0.51$, and $0.59$, respectively.  These are all greater than Geller's estimate of $f_T = 0.34$ \citep{Geller21}, and constitute a basis for concern about the analysis presented here. 

An additional source of information about $f_T$ values in open star clusters is given by \cite{Childs24} and \cite{Childs25}.  This study differed in an important way from \cite{Geller15} and \cite{Geller21} in that Childs and Geller determined binarity on the basis of photometric binaries.  Since photometric binaries are most confidently detected when $q \rightarrow 1$, where $q$ is the mass ratio, \cite{Childs24} and \cite{Childs25} give a parameter $f_{b,q,M}$, defined as the fraction of stars that are binaries, with a primary mass $M$ of roughly 1 solar mass, and a minimum $q=0.5$ (photometric binaries are nearly undetectable otherwise).  

Table A1 of  \cite{Childs25} gives $f_{b,q,M}$ values for a sample of open clusters, which range from a low value of 0.11 to a high of about 0.35.  These numbers should be multiplied by 2 to obtain the binary fraction for all $q$ values, and indicates that the equivalent to my $f_T$ parameter would be $(0.22 \leq A \leq 0.70)$. 

To summarize the results of this subsection, the values of $A$ that I retrieve from the M67 data are high compared with the estimate of \cite{Geller15} and \cite{Geller21}, but not completely deviant given the results of \cite{Childs24} and \cite{Childs25} on a sample of open clusters.  

\subsection{5.2 The Core Dispersion Parameter $\sigma$}
The parameter $\sigma$ is of greatest interest to this study, because it contains information on the amplitude of putative luminosity variations in solar-type stars.  I again emphasize the point made in Section 1.3 above; the value of $\sigma$ retrieved in this analysis will be the quadratic sum of the variances of all processes causing dispersion in the ordinate of a color-magnitude diagram, not just luminosity variations of the primary and secondary, as assumed in the development of \cite{Spangler25a}. To obtain an estimate of the magnitude of true luminosity variations, it is necessary to subtract the variances of the other processes enuntiated in Section 1.3 above.  In the present paper, I consider only the effect of noise errors in the photometric measurements \citep{Spangler25c}, for reasons which will become clear. 

\subsubsection{5.2.1 Dispersion Due to Photometric Error}  
In \cite{Spangler25c}, I considered the effect of photometric errors in the V and B magnitudes (or any other photometric system) on the width of the main sequence in a star cluster.   \cite{Spangler25c} gives the following formula for the Gaussian dispersion ($x$ or $\Delta m$ in the terminology of this paper)
\begin{equation}
\tilde{\sigma} = \sqrt{\frac{A'}{R^2-1}} \sigma_B
\end{equation}
The variables in Equation (7) are defined in \cite{Spangler25c}, and repeated here as  $\sigma_B$,
the rms photometric error in the B band, $A' := 1+R^2a^2+2a$, where $a$ is the slope of the linear approximation to the main sequence in a star cluster, and
 $R^2 := 1+\frac{\sigma_B^2}{\sigma_V^2}$, $\sigma_V$ being the rms photometric error in the V band magnitude. 
 
 Calculating the noise error $\tilde{\sigma}$ requires information on the model isochrone as well as the errors in the magnitude measurements in the data set.  To determine the slope of an approximate linear representation of the main sequence in the region of interest, I did a linear regression to the theoretical isochrone model points (solid red data points in Figure 1), over the color range of interest, $0.60 \leq (B-V) \leq 0.90)$, 
\begin{equation}
m_V(\mbox{isochrone}) = a(B-V)+b
\end{equation} 
  This calculation was done using the Python program Numpy.Polyfit, with polynomial order 1. The returned slope and intercept were $a=+6.20$  and  $b=10.27$.  A comparison of the linear approximation to the theoretical main sequence with the model itself is shown in Figure 6.  In this case, the HR diagram is plotted with the ordinate and abscissa ordered according to normal algebraic usage rather than the convention in astronomy. The true isochrone shows discernible departures from the linear model, but the linear approximation is probably adequate for its intended purpose, which is to extract parameters that make Equation (7) an acceptable approximation to the true noise budget.   
  
\begin{figure}[h]
\begin{center}
\includegraphics[scale=0.50,angle=0]{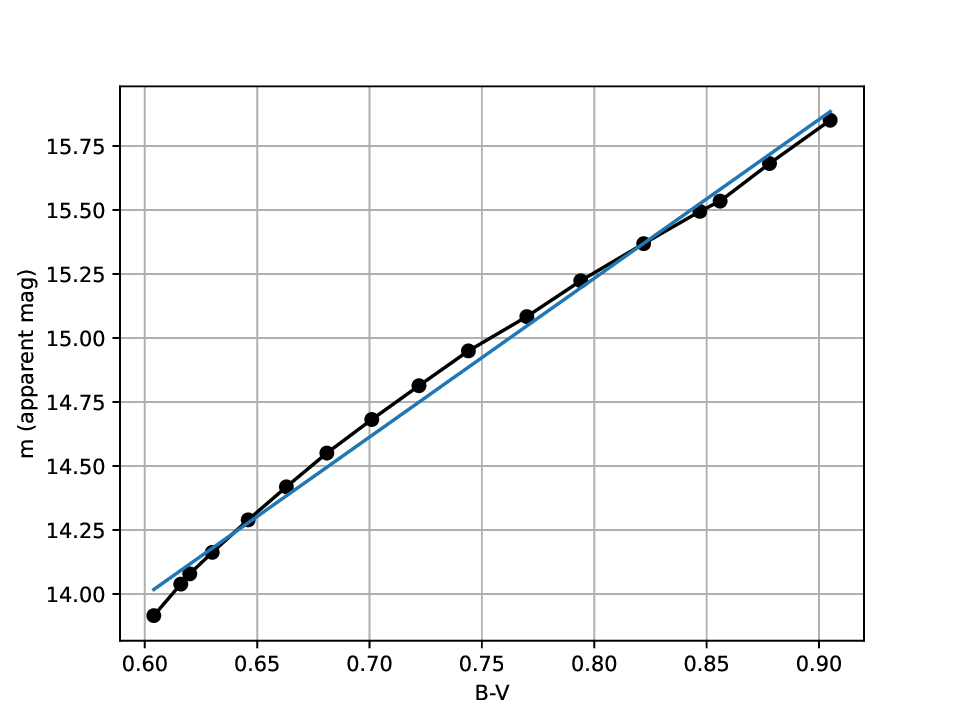}
\caption{Comparison of theoretical isochrone (black solid data points) versus linear approximation (blue line).  The slope and intercept of the line are a=6.20 and b=10.27, respectively.  The linear approximation is used in my expression for the Gaussian width due to photometric error (Equation (7). }
\end{center}
\end{figure}
Information on the photometric errors in the U, B, and V magnitudes for the \cite{Montgomery93} data set are given in Table 6 and Figure 1 of \cite{Montgomery93}.  From this information, I choose $\sigma_B = 0.011$ and $\sigma_V = 0.012$ as {\em characteristic} of the errors for this part of the main sequence in this data set. These values are also consistent with the last sentence in Section 2.3 of \cite{Montgomery93}.  Referring to the dispersion of differences between stellar magnitudes in different photometric studies, \cite{Montgomery93} state ``\ldots the standard deviations are those expected for adding independent data sets with standard deviations of the order of 0.01 mag''.   

With these choices, $\tilde{\sigma} = 0.110$.  Before comparing $\tilde{\sigma}$ to the fit values of $\sigma$ in Section 4, it should be recognized that $\sigma$ describes presumed Gaussian fluctuations in the luminosity of a star, while $\tilde{\sigma}$ measures the approximately Gaussian fluctuations in the magnitude of that object.  There is a logarithmic relationship between these two quantities that needs to be considered.  

It may be shown that if $r$ represents the normalized luminosity of a star, and $r$ is Gaussian-distributed about a mean value of unity with standard deviation $\sigma$, then the pdf of the magnitude fluctuations is approximately Gaussian (for $\sigma \ll 1$) with a mean of zero and a standard deviation of $\tilde{\sigma} = 1.086 \sigma_{\phi}$. What this means is that a fit of a theoretical $p_x(x)$  function from \cite{Spangler25a} to a star cluster with only single stars, and for which the only process causing main sequence broadening is photometric noise, the retrieved parameters $A$ and $\sigma$ (Section 4) would be $A=0$ and $\sigma = \sigma_{\phi}=\tilde{\sigma}/1.086$.

These considerations mean that if intrinsic luminosity variations were entirely absent, the expected photometric errors would produce a retrieved value of $\sigma_{\phi} = 0.101$.  This number is obviously completely consistent with the values of $\sigma$ given in the analysis of Section 4 and Table 1, and suggests that photometric errors dominate any intrinsic variations, and probably  differential extinction and distance effects as well. 
 
\subsection{5.3 Upper Limits to Intrinsic Variations of Solar Type Stars in M67}

Comparison of the photometric-noise-equivalent value of $\sigma_{\phi}$ (the photometric-noise-equivalent value of $\sigma$) calculated above to the results in Table 1 make it clear the data for M67 are consistent with photometric noise being responsible for all or nearly all of the variations in the central core of the $p_x(x)$ distribution shown in Figure 2. The fit to the inner core of the distribution, $-0.30 \leq x \leq 0.30$ with a linear trend removed (Section 4.2) has even less room for any processes other than photometric noise.  

If the range in $\sigma$ resulting from fits to the full empirical $p_x(x)$ distribution is taken (middle row of Table 1), there is a small range of viability for contributions from intrinsic variations.  In what follows, I calculate this contribution, with the goal of determining what upper limit can be placed on intrinsic variations. Let $\sigma$ be the value retrieved from a fit such as that shown in Figure 2.  This value of $\sigma$ contains contributions from the sought-after intrinsic variations (characterized by $\sigma_{IV}$), and photometric variations, described by $\tilde{\sigma}=1.086 \sigma_{\phi}$.  The value of $\sigma$ retrieved in the fit is then
\begin{eqnarray}
\sigma = \sqrt{\sigma_{\phi}^2 + \sigma_{IV}^2} \\ \nonumber
\sigma_{IV} \leq \sqrt{\sigma^2 - \sigma_{\phi}^2}
\end{eqnarray}

If we choose the maximum acceptable value for the full fit from Table 1, $\sigma = 0.135$, and $\sigma_{\phi} = 0.101$, the inferred value of intrinsic variability is $\sigma_{IV} \leq 0.089$ where the $\leq$ symbol indicates that this number is to be understood as an upper limit to any possible intrinsic variation. Use of the best-fit value of $\sigma$ from Table 1, $\sigma = 0.117$, results in an even lower limit to intrinsic variability of $\sigma_{IV} \leq 0.058$. 

The upper limit quoted here includes any contribution of differential extinction across the face of M67, as discussed in Section 1.3 above.  I do not make an estimate of this in the present paper, because measurements of the sort described by \cite{Curtis17} would have to be made on a star-by-star basis for the whole sample, or at least a subset that would permit inference of the statistics of differential extinction. I note that Figure 16 of \cite{Childs24} appears to show $E(B-V)$ varying from $0 \leq E(B-V) \leq 0.05$ over a field perhaps $4^{\circ}$ in diameter, so the contribution might be significant. Pending future investigation of this point, all of this possible upper limit of $\sigma_{IV} \leq 0.058-0.089$ could conceivably be attributed to differential extinction in front of the star cluster.  

\section{6. Are the Limits to Long Term Variability Significant?}
Taking as the main result of this paper an upper limit (possibly generous) of $0.06 \leq \sigma_{IV} \leq 0.09$, one may ask if this result is of any interest or significance.  Although an unambiguous detection of long-term variability would be of great significance to stellar astrophysics, even an upper limit may be of geological and climatological interest.  

\cite{Judge20} analysed photometry of 22 solar-type stars over a 17 year period, and found that some showed secular variations amounting to 0.019 magnitudes over a period of 250 years, the time since the Maunder Minimum.  In terms of ``climate forcing'', this corresponds to a change of 4.5 W/m$^2$ over that time.  \cite{Judge20} note that estimates of anthropogenic climate forcing over that time period are in the range of 1.1 - 3.3 W/m$^2$. A corroborating number is from \cite{Romps22}, who state that doubling of atmospheric $CO_2$ produces an additional radiative forcing of about 4 W/m$^2$.  These changes in the radiative forcing are to be compared with a total solar radiative forcing of 239 W/m$^2$ \footnote{This number comes from taking the total solar irradiance of 1367 W/m$^2$, dividing by 4 to account for the average flux over the entire planet, and multiplying by the complement of an effective albedo of 0.30}. 

A summary of the aforementioned numbers is that the same change in radiative forcing caused by doubling of atmospheric CO$_2$ would be produced by a change in solar luminosity of $\simeq 0.017 - 0.019$. This number is substantially smaller than, but of the same order of magnitude as the upper limit I report here. 
The result from the present study is not yet at the same level that permitted \cite{Judge20} to make the following intriguing statement: {\em ``The stars are therefore tantalizingly close to providing useful constraints on magnetically induced solar irradiance variations, independent of any other measurements or assumptions''}. However, as will be discussed in the next section, substantial improvements in this type of analysis could be made with straightforward advances in data analysis and modeling methods. Nonetheless, the upper limits I report here may already be marginally relevant for assessing the possibility that solar luminosity changes could have been responsible for some of the larger climate variations alluded to in the Introduction. 

An extreme case for comparison is the ``Faint Young Sun Paradox'', which refers to an apparent contradiction between astrophysical and geological conclusions regarding conditions on the early Earth.  A highly readable, and relatively recent discussion of this topic is \cite{Spencer19}.  Figure 2 of \cite{Spencer19} shows that stellar evolution models indicate that 2.5 - 3.2 Gyr ago, the solar luminosity should have been $0.77 - 0.85 L_{\odot}$, where $L_{\odot}$ is the current solar luminosity.  The point of the Faint Young Sun Paradox, of course, is that this conclusion appears to be flatly contradicted by field geological results on the conditions at the Earth's surface during this period.  \cite{Spencer19} offers an intriguing resolution of this paradox.   

The relevance for my present study, however, is that the Faint Young Sun Paradox provides a benchmark for speculated extreme solar luminosity variations during the Phaneozooic Eon.  That variation is substantially larger than the upper limits I quote here in Sections 4 and 5 of the paper.  

\section{7. Possible Future Improvements in this Method} 

Limiting long-term solar luminosity variations from open star cluster observations is especially interesting, because substantial improvements could readily be made to the analysis presented here.  These improvements could be made both in the data set utilized as well as methods of analysis.  
\subsection{7.1 Improvements in Data: Gaia DR3 }
The main conclusion of this whole program of research, contained in \cite{Spangler25a,Spangler25c} as well as the present paper, is that the width of the ``core component'' of the main sequence width (width in the $p_x(x)$ function), is probably dominated by photometric errors.  In this study, I have utilized the measurements of \cite{Montgomery93}, because those were used by and left in machine-readable form by \cite{Geller15}, together with flags indicating known binaries.  The measurements of \cite{Montgomery93} were made with a relatively modest, ground-based telescope.  

The obvious data set to use for a future implementation of this method is Gaia photometry from Gaia DR3 (or later, if available).  Gaia observed the field of M67, which is probably uniquely suited for this method.  In fact, Gaia data for M67 have already been extensively used by \cite{Childs24}. Figure 16 of \cite{Childs24} shows the HR diagram of M67 using Gaia data, and the width of the main sequence is visibly narrower than that in Figure 1 of this paper.  \cite{Childs24} have also applied corrections to the data from a model for differential reddening and extinction, and distinguished between single stars and photometric binaries.  For the purposes of the present paper, use of photometric binaries is questionable because there could be an ambiguity between a true photometric binary and a single star subject to long term luminosity variations.  Nonetheless, a future analysis could proceed with caution, and use the photometric binaries or not.  

\subsection{7.2 Improvements in Modeling: Use of Simulations}
The analysis in this paper has modeled the empirical $p_x(x)$ function for M67 (data points in Figure 2) with largely or exclusively analytic equations presented in \cite{Spangler25a} and \cite{Spangler25b}.  An analytic approach has the advantage of providing insight as well as intellectual gratification\footnote{I feel I prefer this approach because, as an adolescent, I watched the TV series {\em Outer Limits}, which defined physicists as individuals who wrote complicated, arcane equations on blackboards.}. However, at least in the case of the results of \cite{Spangler25a,Spangler25b}, these expressions employed approximations which often elide known properties of stars in isolation and in star clusters.  

An alternative approach, less intellectually satisfying but potentially more accurate, is to simulate the assumed astronomical and mathematical processes.  Monte Carlo simulations were used in the development of \cite{Spangler25a} and \cite{Spangler25c}, but in the restricted role of verifying analytic expressions.  What could be done instead is to simulate all of the processes and stellar properties assumed in the development to this point.  The simulated distribution functions $p_x(x)$ could then be compared with the observed one, and the best model chosen.  In some sense, this may be considered the forward problem corresponding to the ``inverse problem'' approach of \cite{Childs24}, in which the observed stellar data for M67 were submitted to the Base-9 Bayesian analysis algorithm to obtain properties of the individual star clusters as well as member stars within each cluster. 

A specific example of a step in my analysis which would benefit from a simulation approach is in the description of the change in light from a binary system with mass ratio $q$.  The approximation employed in my analysis utilizes Equation (2) of \cite{Spangler25a} (see details in \cite{Spangler25b}), which has the additional light contributed by the binary given by a simple algebraic dependence on the mass ratio $q$.  This essentially has the light from a binary displaced in a vertical direction on a color-magnitude diagram.  While absolutely correct for $q=0$ and $q=1$, this approximation is not correct for intermediate values of $q$, as discussed and illustrated by \cite{Hurley98}.  Although awkward to implement in a largely analytic approach as I have followed, the binary trajectories on a color-magnitude diagram presented in \cite{Hurley98} would be easy to implement in a simulation analysis.  

In conclusion then, the hope is that with a combination of Gaia photometry and a full simulation analysis guided by the analytic approach adopted in this project, upper limits or actual detections of long term variability of solar-type stars would be established which are of interest in a geological, climatological, and astrophysical  context. 

\section{8. Conclusions}  
\begin{enumerate}
\item I have obtained the empirical pdf for the width of the main sequence of the open star cluster M67, utilizing data from \cite{Montgomery93} and \cite{Geller15}, and a theoretical isochrone with the nominal properties of M67.  
\item The \cite{Geller15} data were used to eliminate a large number of binary stars, which will have magnitudes brighter than a theoretical main sequence for single stars, and thus be confused with stars that are  undergoing intrinsic variations.  
\item The width of the main sequence is measured by a variable $x := \Delta m = m_0 - m$, and defined in Equation (1).  The variable $x$ is the difference between the magnitude of a star and the theoretical isochrone magnitude for a star with the same measured color, such as $(B-V)$.
\item A semi-analytic expression for the probability density function $p_x(x)$ \citep{Spangler25a} was fit to the empirical function.  This model  $p_x(x)$ is determined by two parameters, $A$ the fraction of stars in the sample which are binaries (after known binaries are removed), and $\sigma$ which is the root-mean-square, normalized luminosity variation of both the primary and secondary stars. The parameter $\sigma$ contains the information of interest to this investigation, which is the possible amplitude of long-term luminosity variations.   
\item A fit of the theoretical $p_x(x)$ function to the empirical function yielded fits which are marginally acceptable, and led to the following ranges of acceptable parameters, $0.25 \leq A \leq 0.45$ with a best fit value of $A=0.353$, and $0.100 \leq \sigma \leq 0.135$, with a best-fit value of $\sigma=0.117$. A restricted fit to the ``inner core'' of the $p_x(x)$ distribution yields a slightly smaller range of acceptable $\sigma$ values. 
\item The retrieved value of $A$ is high, given that the sample had a large number of known, spectroscopic binaries removed, and is larger than would be expected on the basis of estimates of \cite{Geller21}.  The value is probably not incompatible with results for six open clusters, including M67, presented by \cite{Childs24}.  
\item The value of $\sigma$ retrieved in the fit is entirely consistent with the value produced by photometric noise in the V and B magnitudes used in the color-magnitude diagram. These errors cause a broadening of the main sequence of a star cluster which mimics intrinsic variability of the stars.  An expression for the broadening of the main sequence of a star cluster by photometric noise was developed by  \citep[][see details of derivation in \cite{Spangler25d}]{Spangler25c} and has been applied here.   
\item The maximum value for the amplitude of an intrinsic Gaussian process describing stellar variablility is $0.058 \leq \sigma_{IV} \leq 0.089$.  This value may be compared with the increase in solar luminosity which would cause radiative forcing equal to that caused by doubling of atmospheric CO$_2$, which is $\simeq 0.017$.  The upper limit obtained from the M67 data is therefore larger, but of the same order of magnitude.  It could be that luminosity variations of this magnitude could be responsible for other, more extreme climate variations in the geological record. 
\item An analysis of this sort using much more precise photometry from the Gaia spacecraft, and more sophisticated modeling methods would yield much lower limits, or possibly a detection of long term variability of solar type stars.  
\section{Acknowledgements}
I thank Dr. Aaron Geller of Northwestern University for sharing his magisterial knowledge of open star clusters, and making numerous helpful suggestions during this research project, and Dr. Kenneth Janes of Boston University for advice on the magnitude of photometric errors in the data set I used.  I also thank Dr. David Peate of the School of Earth, Environment, and Sustainability at the University of Iowa, for bringing to my attention the article by J. Spencer on the Faint Young Sun Paradox.  
\end{enumerate}

\end{document}